\documentclass[11pt]{article}
\usepackage{graphicx}

\newcommand{\beq}{\begin{equation}}
\newcommand{\eeq}{\end{equation}}
\newcommand{\beqa}{\begin{eqnarray}}
\newcommand{\eeqa}{\end{eqnarray}}

\newcommand{\aj}{{\it Astron. J.  }}
\newcommand{\apj}{{\it Astrophys. J. }}
\newcommand{\aap}{{\it Astron. Astrophys.}}
\newcommand{\mnras}{{\it Mon. Not.  R. Astron. Soc. }}
\newcommand{\apjl}{{\it Astrophys. J. Letters }} 
\newcommand{\apjs}{{\it Astrophys. J. Supplement }}
\newcommand{\nat}{{\it Nature }}
\newcommand{\prd}{{\it Phys. Rev. D }}
\def\araa{{\it Ann. Rev. Astron.  Astrophys. }}

\def\la{\lower.5ex\hbox{$\; \buildrel < \over \sim \;$}}
\def\ga{\lower.5ex\hbox{$\; \buildrel > \over \sim \;$}}

\begin{document}

\noindent{\bf Clues from nearby galaxies to a better theory of cosmic evolution }

\bigskip
\noindent{P.J.E. Peebles\footnote{Joseph Henry Laboratories,
    Princeton University, Princeton, NJ 08544, USA} and Adi Nusser\footnote{Physics department, Technion, Haifa 32000, Israel} }
    
\bigskip

\noindent {\bf The great advances in the network of cosmological tests show that the relativistic Big Bang theory is a good description of our expanding universe. But the properties of nearby galaxies that can be observed in greatest detail suggest a still better theory would more rapidly gather matter into galaxies and groups of galaxies. This happens in theoretical ideas now under discussion.}
\bigskip

\noindent In the standard cosmology \cite{bigbang} 4\% of the mass of the universe is in the baryons of which stars and planets are made, 22\% is nonbaryonic dark matter (DM), and the rest is Einstein's cosmological constant (or acts like it). Gravity has collected the dark matter in concentrations termed DM halos. In larger DM halos baryons were dense enough to have radiated away enough energy to collapse to galaxies and stars. The most massive halos, natural homes for the most luminous galaxies, preferentially form in regions of the highest local mass density. Less massive halos, natural homes for less luminous galaxies, appear in regions extending to  lower local densities, in ridges of matter running between denser regions, forming a cosmic web \cite{cosmicweb} of filaments and sheets (as in Figs.~1 and 2 in \cite{MathisWhite}). This can be compared to the situation in our immediate extragalactic neighborhood illustrated in Figure~1. The Local Sheet at SGZ = 0 certainly looks like part of the cosmic web of the standard cosmology, but there are problems. 

Our selection of observations that seem to be pointing to an improved theory commences with the least crowded place in Figure~1, the Local Void, which contains far fewer galaxies than expected, while there is an unexpected presence of large galaxies on the outskirts of the Local Void. These problems would be eased if structure grew more rapidly than in the standard theory, more completely emptying the Local Void and piling up matter on its outskirts. In about half of the largest galaxies in Figure~1 the stars are largely confined to thin disks supported by streaming motions in the plane of the disk. These pure disk galaxies do not appear in numerical simulations of galaxy formation in the standard theory, because the relatively slow assembly allows stars to accumulate in thick stellar bulges or halos. Again, the problem would be relieved by more rapid structure formation, with an earlier termination of the rain of extragalactic debris onto galaxies. Observations at greater distances of bigger samples of galaxies show that the general properties of large galaxies in the two great classes, spiral and elliptical, have strikingly little relation to their surroundings. Yet again, this would be a more natural outcome of a theory that predicted earlier formation and isolation from the surroundings.  

We have chosen these issues among many open questions in astronomy because they seem to be least likely to be affected by the complexities of interactions of stars and gas. It is the variety of this collection of apparently serious challenges to the standard theory that makes the case for considering a still better theory of cosmic evolution in the expanding universe.

\begin{figure}[htpb]
\begin{center}
\includegraphics[angle=0,width=5.in]{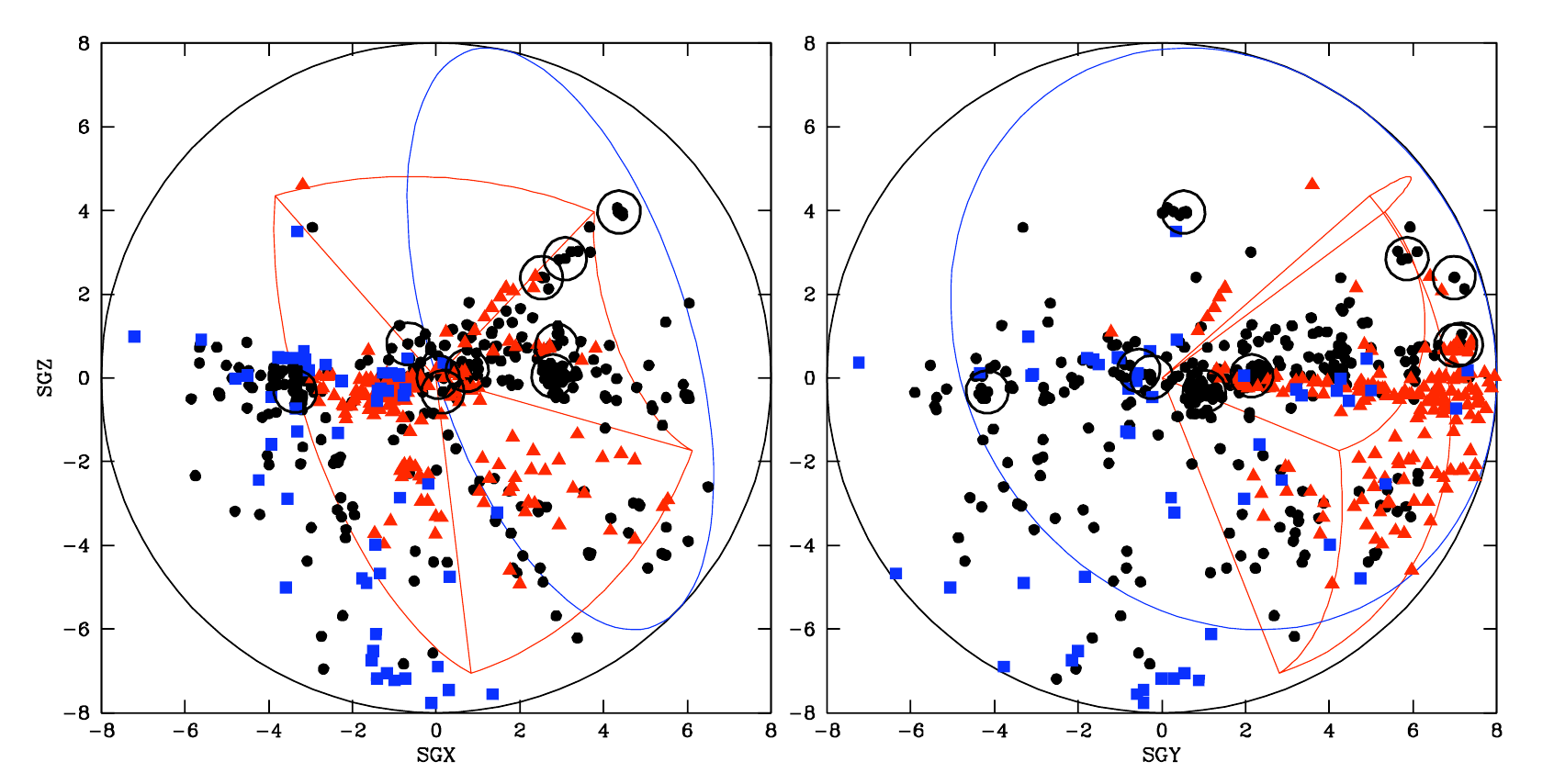} 
\caption{Galaxies at distances $1<D<8$\,Mpc from the center of the Local Group of galaxies. The Local Sheet is the concentration along the mid plane, the Local Void the region to the upper left in the left-hand projection. The ten most luminous galaxies (including M31 and the Milky Way at $D<1$\,Mpc) are at the circles. The orthogonal projections are plotted in supergalactic coordinates \cite{deV53}. Black circles: 337 galaxies  largely discovered on photographic plates and with secure distance measurements \cite{Kara}. Red triangles: 172 galaxies added by the digital SDSS survey \cite{SDSS} with redshift errors less than 50 km s$^{-1}$. Blue squares: 53 galaxies discovered by the HIPASS survey for 21-cm emission by atomic hydrogen \cite{HIPASS}. SDSS and HIPASS have less secure redshift distances, and cover only the parts of the sky roughly indicated by the red and blue curves. There are many more dwarfs to be discovered at this distance. \label{fig:1}}
\end{center}
\end{figure}

\begin{figure}[htpb]
\begin{center}
\includegraphics[angle=0,width=3.in]{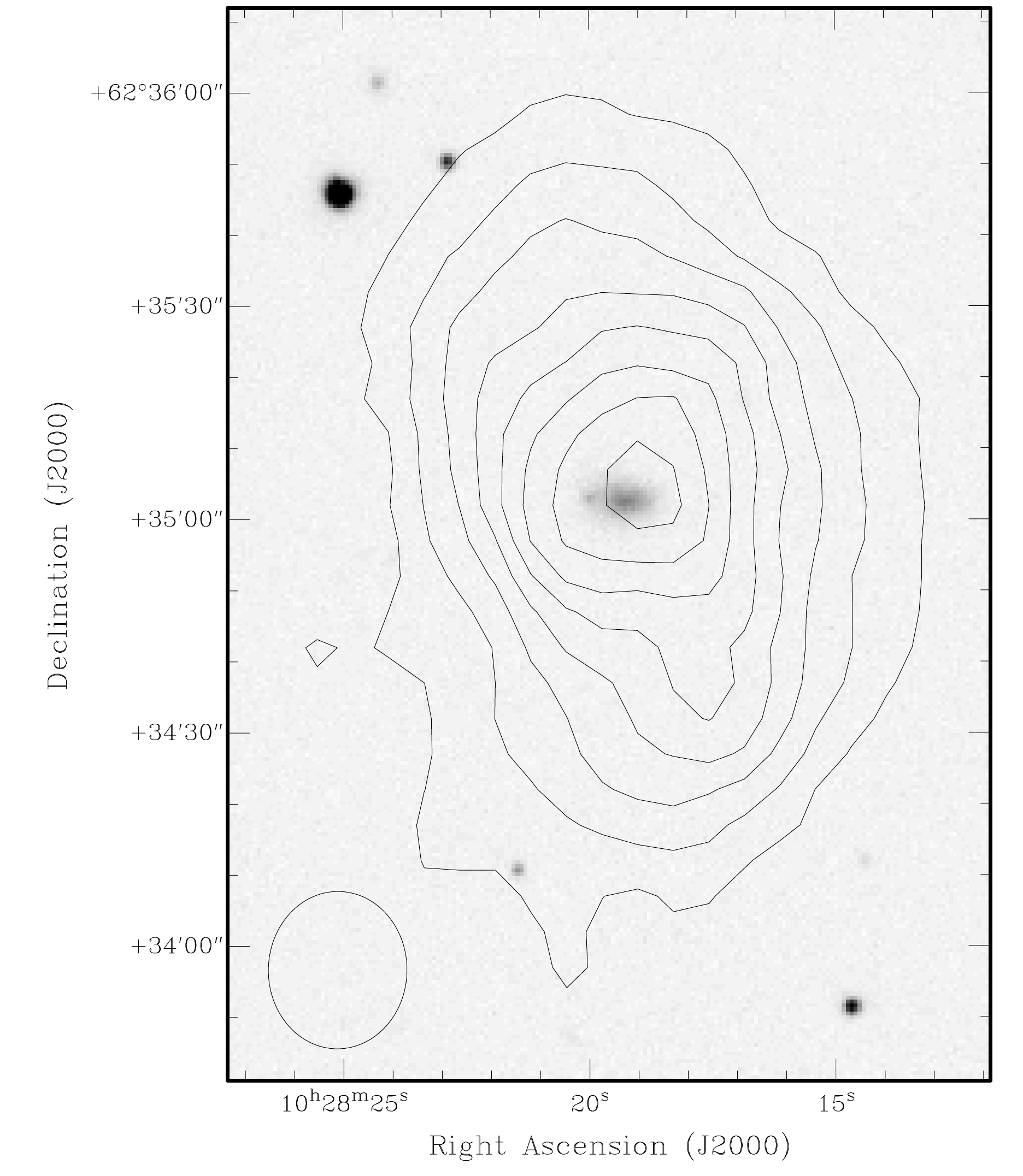} 
\caption{A galaxy typical of those found in low density regions. The contours trace the extended rotationally-supported disk of atomic hydrogen. The central gray patch is a negative image of the more concentrated starlight.\label{fig:2}}
\end{center}
\end{figure}

\bigskip
\noindent{\bf Galaxies in the Local Void} \medskip

\noindent Galaxies are less common in low density regions than expected in the standard cosmology \cite{Peebles07, Klypin09}. The Local Void (the LV)  in Figure~1 contains just three of the 562 known galaxies in this distance range. The left-hand projection shows the three centered on the LV; the orthogonal projection shows they are well separated. Two are detected in 21-cm emission from their atomic hydrogen. Images of other galaxies in quite uncrowded regions \cite{Stanonik} suggest that these two LV galaxies have a  disk of atomic hydrogen that extends well beyond the stars, as in the example in Figure~2. (The LV galaxy shown as the red triangle is not detected in 21-cm emission. It may be an exception to the rule that galaxies in  uncrowded regions have abundant hydrogen, or perhaps the distance error wrongly put it in the LV.)
  
Numerical simulations predict the number density of small DM halos in a low mass density region is 10\% of the global mean, the halos appearing more commonly near the edges  \cite{Gottlber03}. The LV occupies about a third of the volume at $1<D<8$\,Mpc. If environment had no effect on the probability a DM halo is visible in starlight or the 21-cm line then these numbers would predict that about 19 of the 562 known galaxies are in the LV. The formal Poisson probability for just three LV galaxies is negligibly small (about 10$^{-5}$), but the issue is systematic errors, as follows

The predicted number of LV galaxies should not be confused with the prediction of many more dwarf DM halos than there are dwarf galaxies \cite{Zwaan}. The latter could mean many small halos have lost too much baryonic matter to be detectable \cite{Ostriker_MoverL, Kravtsov}. A recent analyses \cite{TinkerConroy} of this effect indicates that low mass halos may have such low luminosities that the standard cosmology would lead to an acceptable number of LV galaxies. But our predicted number, 19, is derived from the count of nearby galaxies that actually are detected. If the predicted number in the LV were significantly lowered it would predict too few observed dwarfs in the rest of Figure~1. This problem would be relieved if the galaxy count at $D<8$\,Mpc were well above the cosmic mean while the mass density in the LV has the average void density. The arrangement is contrived, but it should be explored. Another very real and potentially serious systematic error is that there are many more galaxies to be discovered and placed in Figure~1. The LV has already been surveyed for the hydrogen-rich galaxies characteristic of low density regions, but future still more sensitive surveys will be of great interest.  

There is a systematic error that likely exacerbates the problem for the standard cosmology. We have assumed a DM halo of given mass has the same chance of being detectable wherever it is. But dwarf galaxies in more crowded regions, closer than about 0.3\,Mpc to a large galaxy, tend not to contain much atomic hydrogen, perhaps because it has been stripped away by the ram pressure of plasma around larger galaxies \cite{vandenBergh, GrcevichPutman(2009)}. The more isolated dwarfs tend to contain detectable atomic hydrogen, as in  the blue squares in Figure~1.  That is, the evidence is that  the apparently tranquil environment in voids in the distribution of large galaxies offers the best chance for survival of low mass DM halos with the apparently fragile disks illustrated in Figure~2. This would mean that the fraction of DM halos that are detectable by radiation from their hydrogen is larger in the LV than elsewhere, so we would have expected more than 19 LV dwarfs, not the three found so far.  

We conclude that there is a good case for inconsistency of the theory and observation of galaxies in the LV. Conceivably the local sample is atypical; that will be checked as galaxy surveys improve. Perhaps survival of detectable galaxies is less likely in the LV, though that is contrary to the depleted state of dwarfs near large galaxies. Or perhaps we are learning that growth of structure is more rapid than predicted in the standard cosmology, more completely emptying low density regions. 
\bigskip

\noindent{\bf Large Galaxies in Underdense Regions} \nopagebreak\medskip

\noindent The ten most luminous nearby galaxies are circled in Figure~1. (The brightest has just twice the luminosity of the tenth brightest.  Most of the rest are much fainter, down to about $10^{-5}$ times the brightest.) The spiral galaxy NGC 6946 is among the ten brightest and among the most  isolated. It and its several satellites are the little island at  the highest circle in Figure 1. Its disk of atomic hydrogen is more than twice the extent of its visible stars \cite{N6946disk}, another example of the tendency of galaxies in less crowded regions to have more extended atomic hydrogen disks. 

Among the ten most luminous galaxies in Figure~1 the spirals M\,51, M\,101, and NGC\,6946 are 2.4, 2.8, and 4.0\,Mpc above the center plane of the Local Sheet. They are in an uncrowded region: of the 562 known galaxies at $1<D<8$ Mpc just 5.0\% are more than 2~Mpc above the plane of the Local Sheet (while 73\% of the known galaxies are within 2~Mpc of the plane and the rest are below the plane). But 30\% of the largest galaxies are more than 2~Mpc above the Local Sheet. If galaxy luminosities were randomly assigned this situation would have a 1\% probability.  But the probability is less than this in the standard picture of the cosmic web where more luminous galaxies avoid less dense regions. These three could not be dwarfs masquerading as large galaxies; their circular velocities indicate the central masses of large galaxies. That is, the presence of these three large galaxies in the uncrowded region above the  Local Sheet is  real, and at well below 1\% probability it is unlikely within standard ideas. 

What might this mean? To be analyzed is the effect of physics that more rapidly empties voids, as wanted to account for the scarcity of galaxies in the LV. That would push small galaxies that formed in the low density inner region out past larger galaxies that formed nearer the higher density edge of what will become the void, forming a ridge \cite{vandeWSheth}, perhaps the Local Sheet. And this process, with more rapid emptying of the LV, may enlighten us on the curious presence of large galaxies above the Local Sheet.

\begin{figure}[htpb]
\begin{center}
\includegraphics[angle=0,width=3.in]{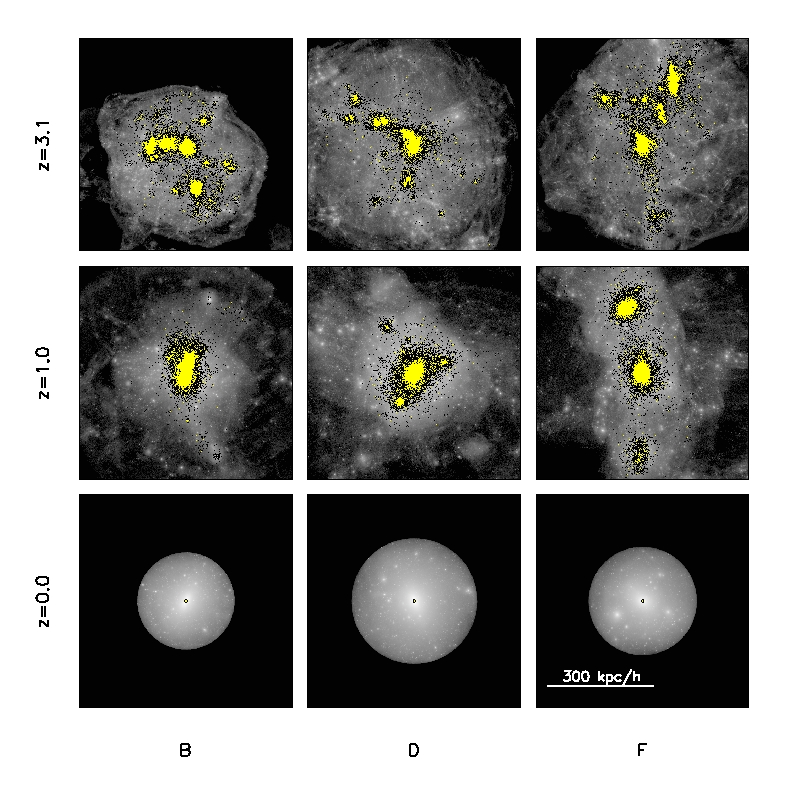} 
\caption{Ongoing rearrangement of matter now in the central luminous regions of galaxies.  In this pure DM simulation the halo masses are suitable homes for spiral galaxies similar to the Milky Way. Jie Wang, Durham University, made the figure from the Acquarius \cite{Springel} simulation. The letters identify the halos at level 2 in the first column of the simulation parameters in Table~1 in \cite{Springel}. Dark matter particles closer than 2.9 kpc from the halo center at the present epoch are plotted in yellow in all panels, particles now at $2.9 < D < 7$  kpc are plotted in black, particles now outside 7~kpc but within the part of the halo that is now nominally close to statistical equilibrium are plotted in grey tones that represent their surface number density, and particles now further away are not plotted in any of the panels. The stars in a galaxy like the Milky Way are largely within about 7 kpc, the outer radius at the present epoch for the particles plotted in black. The box width, 830~kpc, is constant in physical units. Present positions (redshift $z=0$) are shown in the bottom row, positions when the universe was half its present size ($z = 1$) are in the middle row, and positions when the universe was one quarter its present size ($z=3$) are in the top row. The matter falling into the central regions that are largely occupied by the stars in a real galaxy tends to be balanced by matter flowing out \cite{Gao_etal}. The rain does not so much add mass to the central parts of a galaxy galaxy as rearrange it. 
\label{fig:3}}
\end{center}
\end{figure}

\bigskip
\noindent{\bf Pure Disk Galaxies} \nopagebreak\medskip
 
\noindent Some spiral galaxies have a central concentration of stars in a bulge that is thicker than the disk and supported largely by random motions of its stars. (The kinetic energy of the streaming motion of rotation of the bulge is a subdominant source of support.) Others do not have a significant bulge; the stars are concentrated in a disk or central bar supported by streaming motions in the plane of disk. There may be a concentration of starlight toward the center of one of these pure disk galaxies, but this ``pseudobulge'' is largely supported by rotation \cite{KormendyFisher, KormendyKennicutt(2004), FalconBarroso}. Bulges are a natural outcome of structure formation in the standard cosmology: ongoing accretion of intergalactic debris tends to place stars in orbits that take them out of the disk \cite{SteinmetzNavarro, Frenk_view}. But this cannot have happened to pure disk galaxies, which  includes about half the largest 20 galaxies in Figure~1 \cite{KormendyFisher}.

The long-lasting rain of debris onto large galaxies predicted by the standard cosmology is illustrated in Figure~3. This simulation follows the motion of dark matter only (though a star once formed moves as a dark matter particle). The simulation indicates that the matter now within the parts occupied by stars in a real galaxy was spread over a region two orders of magnitude broader at redshift $z=1$.  (The universe has expanded by the factor $1+z=2$ since then.) Star formation was at its peak from then back to $z=3$, and at least half of all stars formed before $z=1$ \cite{OstrikerFukugita}. The stars that formed  at $z>1$ in the broadly scattered clumps of matter in the figure would end up in a galaxy today. But that is unacceptable for a pure disk galaxy; how could this rain of stars have ended up moving in a disk?   

There are informative results from simulations that ignore the infalling stars and focus instead on the behavior of the infalling gas and plasma. Careful attention to the star formation model can lead to simulated galaxies in which the diffuse matter settles into stars in a suitably extended disk \cite{Hopkins}, and to simulated dwarf galaxies that have realistic distributions of dark matter, stars and gas \cite{Governato}. But simulations \cite{Scannapieco1} of galaxies with mass similar to that of the Milky Way indicate that rotationally-supported disks form only outside 2 kpc radius, which is very different from nearby pure disk galaxies in which rotational support is observed to extend inward of 0.2 kpc. These simulated large galaxies may have bulges that are modest enough to avoid detection when observed at ten times the distance of Figure~1 \cite{Scannapieco2}, but at the smaller distances it is possible to see the details that show that the present art of simulations does not produce good approximations to pure disk galaxies, even ignoring infalling stars.

Of course, we cannot ignore the stars that would have fallen into the galaxies along with the gas and plasma. A pure disk galaxy has to have formed out of an isolated cloud of diffuse matter that was assembled before appreciable numbers of stars had formed, for otherwise there would be unacceptable numbers of stars outside the disk. Numerical simulations show that such a cloud of  diffuse matter without stars could collapse to a pure disk of stars \cite{Hopkins,Governato}. But since the bulk of the stars had formed by redshift $z=1$ the matter going into a pure disk had to have been assembled as a gaseous cloud well before $z=1$. This agrees with the dating of stars in the Milky Way, which is classified as a near pure disk: recent analyses \cite{Wyse09} continue to support the conclusion that if accretion and minor mergers promoted stars from the thin to thick disk component, or added stars to the disk, then this activity ceased at redshift about $z= 2$. 

The conditions for formation of pure disk galaxies out of isolated clouds do not agree with the picture for the evolution of cosmic structure indicated by Figure~3, in which extensive infall of matter continues past redshift $z=1$. How could stars in the rain of material shown in this figure have arranged themselves into pure disks? How could extended rotationally-supported disks of atomic hydrogen, as in the pure disk galaxy NGC\,6946 \cite{N6946disk}, have survived the buffeting by gas in the rain of debris predicted by the standard cosmology?  The way forward certainly will include detailed analyses of the effects on the diffuse gas and plasma by stellar formation, winds and explosions, as in current analyses of galaxy formation in the standard cosmology. But we are led to repeat an earlier theme; a successful theory certainly would be aided by more rapid assembly of the pieces of a galaxy.

\begin{figure}[htpb]
\begin{center}
\includegraphics[angle=0,width=3.in]{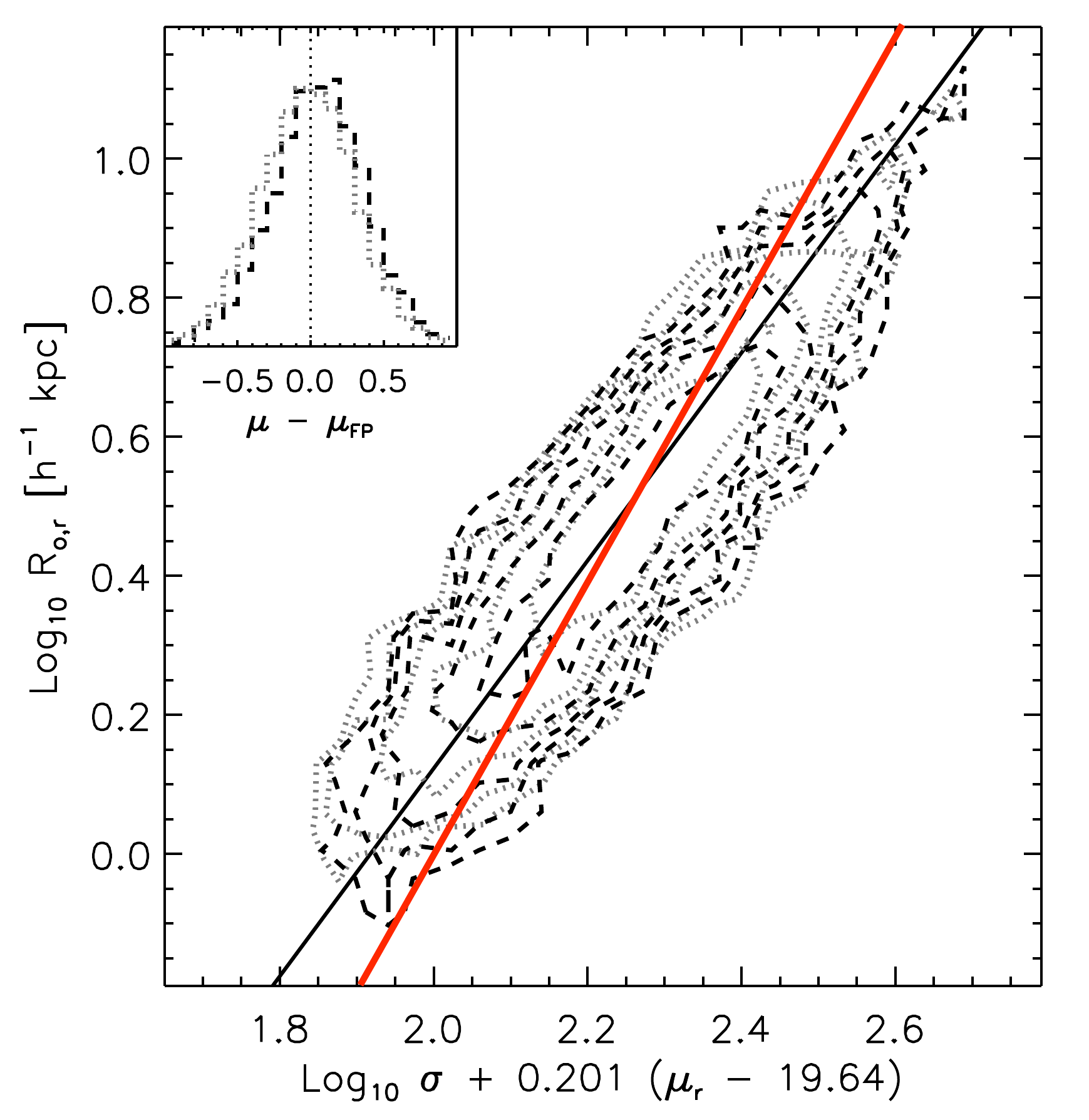} 
\caption{Measures of early-type galaxies in more and less crowded environments. The vertical axis shows the galaxy radius $R$; the horizontal axis shows a product of the galaxy radius, luminosity $L$, and line-of-sight star velocity dispersion $\sigma$ (where $\mu$ is proportional to the logarithm of the surface brightness $\propto L/R^2$). If the distributions of mass and star velocities were similar in these galaxies the condition of balance of gravity against the pressure of the star velocities would require that the mass varies as $M\propto R\sigma^2$, and if the mass $M$ were proportional to the luminosity $L$ the variables would follow the red line. The contours showing the distributions of measurements are tilted  to $M/L\propto R^{0.3}$; larger galaxies have larger apparent mass-to-light ratios. The dashed and dotted contours show there is a systematic difference between $M/L$ at given $R$ in galaxies in more and less crowded regions, but the difference is small. \label{fig:4}}
\end{center}
\end{figure}

\bigskip
\noindent{\bf Galaxies as Island Universes} \nopagebreak\medskip

\noindent  Measures of the sizes, stellar masses, and stellar populations  of large samples of galaxies indicate that these general properties are quite insensitive to the abundance of neighboring galaxies. This also seems contrary to the broad-scale mixing of the contents of present-day galaxies indicated in Figure~3.

Statistical samples of galaxies are conveniently separated into late and early types. Most of the brighter galaxies in Figure~1 are late-type spirals. The early type includes ellipticals that resemble classical bulges with at most small disks and S0 galaxies that bring to mind spirals that have long since lost their gas and with it young stars. Larger early-type galaxies have larger mass per unit of starlight. This correlation, shown in Figure~4 \cite{Bernardi}, is tight, but the main point is that the systematic difference between the correlation for galaxies in more and less crowded environments is exquisitely small. Other correlations that are tight enough to allow little room for sensitivity to environment have been known for a long time \cite{FaberJackson, TullyFisher, VisvanathanSandage, Kormendy77, Szomoru96,  Sage97}. The new development is much larger samples with tighter control of systematic errors made possible by digital detectors and data handling. Recent examples of insensitivity to environment include the relation between luminosity and color, which depends on the ages of the stars and their heavy element abundances \cite{Hogg2004, Vogeley2007}, the relation between luminosity and radius \cite{vandenBergh_etal}, and the relation among luminosity, radius, and velocity dispersion \cite{Bernardi, Disney}; see also the review in \cite{BlantonMoustakas}. The less detailed observations allowed by these  measurements of large samples of galaxies obscure distinctions of essential features such as classical bulges and pseudobulges. But the great advantage is that large samples allow sensitive statistical measures capable of showing real but exceedingly small systematic effects of environment on galaxies.

Environment certainly plays a role in galaxy formation, but these observations tell us it has to be a subdominant factor in determining the properties of galaxies of given type. The brightest galaxy in a rich cluster tends to be exceptionally luminous and extended \cite{TremaineRichstone, Lin_etalBCG}, a likely result of interactions among cluster members \cite{OstrikerTremaine1975}, but one that has affected a minority of early-type galaxies. Some galaxies outside clusters clearly have been disturbed by close encounters with neighbors; examples are in Figure 16 in \cite{BlantonMoustakas}, with an estimate that manifest disturbances appear in 1 to 2 percent of larger galaxies. What will be the natures of disturbed systems after relaxation to a closer to steady state? The disturbance would promote conversion of gas to stars and change rotational streaming of stars in disks to random motions, together producing a system similar to a bulge or an elliptical galaxy \cite{Toomre77, Schweizer82}. But merging of a pair or group of spirals  in the universe as it is now  would produce a new disk that is too young for a standard spiral, or a new elliptical that does not have the standard pattern of stellar ages and heavy element abundances \cite{Ostriker09}. Recent mergers have happened, and have produced odd objects, but the systematics, as in Figure~4, indicate odd galaxies are rare. 

Another aspect of environment is that regions that are significantly more crowded than the cosmic mean host a larger fraction of more luminous \cite{vandenBergh, Hogg2004} and early types \cite{PostmanGeller84} of galaxies. This agrees with the standard picture of formation by merging, but there are conditions. To preserve the color-luminosity correlation the stars entering a protoelliptical have to have the colors appropriate to the elliptical they are going to form. To preserve the radius-luminosity relation the amount of energy carried into a merger, less what is dissipated, has to have been well correlated with the final product. And  preserving the relation in Figure~4 requires that the matter entering a galaxy ``knows'' the effective mass-to-light ratio belonging to what is going to be the mass of the final product. This would be more comprehensible if the parts of a protogalaxy were assembled in conditions well correlated with the final galaxy mass before there was abundant star formation. For example, the tails of stars on the outskirts of nearby ellipticals \cite{vanDokkum} seem to be products of merging, but of stars close enough to the host galaxy that one can imagine stars in host and tails formed in close enough proximity that conditions were similar. How could the large-scale rearrangement of material entering a galaxy shown in Figure~3 and in \cite{Gao_etal} be reconciled with the exquisitely small effect of environment in  Figure~4?

There is a related issue. Nearby large spirals have acquired quite similar rotation in regions that now are quite different; compare the isolation of NGC\,6946 in Figure~1 with the crowded Milky Way in the Local Sheet. If tidal torques from neighboring protogalaxies transferred similar angular momenta to these two galaxies then the two protogalaxies would have to have been in similar environments at formation. To be analyzed is whether the considerable difference in environments now and the required similarity of environments when the galaxies formed can be reconciled with the slow emptying of low density regions, as around NGC\,6946, by standard gravity. 

We conclude with assessments of these correlations by two groups of observers. Nair, van den Bergh, and Abraham \cite{vandenBergh_etal}: ``It is not clear why objects that might have been assembled in such very different ways, from different ancestral objects, should have had evolutionary tracks that converged to show small dispersions around simple power law forms for their size-luminosity relations.'' Disney {\it et al.} \cite{Disney}: ``Such a degree of organization appears to be at odds with hierarchical galaxy formation, a central tenet of the cold dark matter model in cosmology.'' In short, the general insensitivity of galaxies to their  environments is not expected in standard ideas. Again, it would help if galaxies were more rapidly assembled so they could then evolve as more nearly isolated island universes.

\bigskip
\noindent{\bf Prospects}\medskip

\noindent  The most distant detectable galaxies are observed as they were when they were young, because of the light travel time \cite{Bezansonetal, Schreiberetal09}. This offers a deeply important window into how the galaxies  formed. Interpretation of what is observed requires the usual assessment of whether apparent conflicts between  theory and observation might be reconciled by more subtle analysis, but here under the condition of much less detailed observations of individual galaxies. The way forward requires balanced support for fascinating observations of the past and for work on the benchmark, the natures of galaxies as they are now. The benchmark certainly should and can be made more secure. For example, the pioneering survey for hydrogen-rich galaxies like the one in Figure~2 that may lurking in the larger Bo\"otes void \cite{Szomoru96} was followed by the HIPASS blind sky survey \cite{HIPASS}, which is now being followed by the Arecibo  radio telescope survey \cite{Alfalfa}. And still deeper advances are technically quite possible and desirable.  

The variety of problems we have considered in the interpretation of the present baseline motivates serious consideration of adjustments of the fundamental theory. Any adjustment would have to preserve the properties of the standard cosmology that agree with the cosmological tests, but that allows new physics operating on the scale of galaxies. The evidence for more rapid structure formation agrees with what happens in modifications of the gravity physics of general relativity theory \cite{DGP00, Chameleon, Trodden}, and in general relativity with a long-range force acting only on the dark matter \cite{FarrarPeebles, Nusseretal05}. These ideas can be tested by observations of the nature and rate of growth of large-scale structure \cite{Zhangetal07, Huietal09}, and by numerical simulations of what happens in these theories on the scale of galaxies \cite {Keselmanetal, HellwingJuszkiewicz, MartinoSheth,KeselmanetalRE}. Both approaches can be considerably elaborated. 

Two clear conclusions from this review are that galaxy formation is not well understood and that the nearby galaxies offer rich and still far from completely explored clues to a better picture of how the galaxies formed. We have mentioned the idea that more rapid structure formation would help interpret the observations; other ideas may emerge as issues raised by the nearby and distant galaxies are more deeply explored and more widely debated. We do not anticipate that this debate will lead to a substantial departure from the present standard picture of cosmic evolution in the hot Big Bang, because the picture passes a tight network of tests. But there is considerable room for adjustment of details, including the galaxies.

\bigskip

\noindent{\bf Acknowledgments}  We are grateful to the Virgo Consortium for their cosmological simulations, and to Jie Wang (Institute for Computational Cosmology, Durham University), who made Figure~3 from these simulations.  We have profited from advice from Fabio Governato, Anatoly Klypin, John Kormendy, Joe Silk, Sidney van den Bergh, Jacqueline van Gorkom, Simon White, and Rosemary Wyse.  This work was supported in part by The Israel Science Foundation (grant No.303/09).

\bigskip

\noindent{\bf Author Contributions} Both authors contributed equally to this work.  
\bigskip

\noindent{\bf Author Information} Correspondence should be addressed to P.J.E. Peebles 

\noindent (pjep@princeton.edu).

\end{document}